\documentclass[twocolumn,superscriptaddress,prl]{revtex4-1}
\usepackage{mathrsfs,braket}
\usepackage{amssymb, amsbsy, amsmath, latexsym, dsfont, array, layout,
graphicx,mathrsfs,color,ulem,bm,extarrows,bm}

\newcommand{\one}{\mathds{1}}

\begin{document}

\title{Simulating dynamic quantum phase transitions in photonic quantum walks}
\author{Kunkun Wang}
\affiliation{Beijing Computational Science Research Center, Beijing 100084, China}
\affiliation{Department of Physics, Southeast University, Nanjing 211189, China}
\author{Xingze Qiu}
\affiliation{Key Laboratory of Quantum Information, University of Science and Technology of China, CAS, Hefei 230026, China}
\affiliation{Synergetic Innovation Center in Quantum Information and Quantum Physics, University of Science and Technology of China, CAS, Hefei 230026, China}
\author{Lei Xiao}
\affiliation{Beijing Computational Science Research Center, Beijing 100084, China}
\affiliation{Department of Physics, Southeast University, Nanjing 211189, China}
\author{Xiang Zhan}
\affiliation{Beijing Computational Science Research Center, Beijing 100084, China}
\affiliation{Department of Physics, Southeast University, Nanjing 211189, China}
\author{Zhihao Bian}
\affiliation{Beijing Computational Science Research Center, Beijing 100084, China}
\affiliation{Department of Physics, Southeast University, Nanjing 211189, China}
\author{Wei Yi}\email{wyiz@ustc.edu.cn}
\affiliation{Key Laboratory of Quantum Information, University of Science and Technology of China, CAS, Hefei 230026, China}
\affiliation{Synergetic Innovation Center in Quantum Information and Quantum Physics, University of Science and Technology of China, CAS, Hefei 230026, China}
\author{Peng Xue}\email{gnep.eux@gmail.com}
\affiliation{Beijing Computational Science Research Center, Beijing 100084, China}
\affiliation{Department of Physics, Southeast University, Nanjing 211189, China}
\affiliation{State Key Laboratory of Precision Spectroscopy, East China Normal University, Shanghai 200062, China}

\begin{abstract}
Signaled by non-analyticities in the time evolution of physical observables, dynamic quantum phase transitions (DQPTs)
emerge in quench dynamics of topological systems and possess an interesting geometric origin captured by dynamic topological order parameters (DTOPs).
In this work, we report the experimental study of DQPTs using discrete-time quantum walks of single photons. We simulate quench dynamics between distinct Floquet topological phases using quantum-walk dynamics, and experimentally characterize DQPTs and the underlying DTOPs through interference-based measurements.
The versatile photonic quantum-walk platform further allows us to experimentally investigate DQPTs for mixed states and in parity-time-symmetric non-unitary dynamics for the first time. Our experiment directly confirms the relation between DQPTs and DTOPs in quench dynamics of a topological system, and opens up the avenue of simulating emergent topological phenomena using discrete-time quantum-walk dynamics.
\end{abstract}

\maketitle

The study of phase transitions lies at the core of the description of equilibrium states of matter~\cite{landau}. Besides conventional continuous phase transitions that are signaled by symmetry breaking, topological phase transitions, characterized by the change of topology in their ground-state wavefunctions, have attracted much attention since the discovery of quantum Hall effects~\cite{HKrmp10,QZrmp11}. Recent experimental progress has further led to the exciting possibility of creating novel quantum phases of matter in dynamical processes~\cite{ETHcoldatom14,Weitenberg2016,Jo17,Weitenberg1709,Weitenberg17,KB+12,Cardano2016,Cardano2017,PTsymm2,Zeunerprl,pxprl,iondqpt,iondqpt2,chenshudqpt,greinerdqpt}, and thus raised the challenging question on the understanding of emergent phases and phase transitions in non-equilibrium dynamics.

Proposed as temporal analogues to continuous phase transitions, dynamical quantum phase transitions (DQPTs) are associated with non-analyticities in the time evolution of physical observables~\cite{Heyl13,Heyl15,Heyl17}, and have been experimentally observed in various systems~\cite{iondqpt,iondqpt2,chenshudqpt,greinerdqpt}.
DQPT occurs as a consequence of the emergence of dynamic Fisher zeros~\cite{fisherbook,loschmidt}, where the Loschmidt amplitude $G(t)=\langle\psi(0)|\psi(t)\rangle$ vanishes at critical times and the corresponding rate function $g(t)=-1/N \ln |G(t)|^2$ becomes non-analytical~\cite{Heyl17}.
Here $|\psi(t)\rangle$ is the time-evolved state, and $N$ is the overall degrees of freedom of the system.
Whereas it is still unclear to what extent key concepts of continuous phase transitions can be extended to describe DQPTs, an intriguing discovery is the geometric origin of DQPTs, captured by dynamic topological order parameters (DTOPs), which suggests the intimate connection between DQPTs and emergent topological phenomena in dynamic processes~\cite{BH16,mixeddtop,dtop2d,wydtop}.

A particularly important scenario is the quench dynamics of topological systems, where the ground state $|\psi^{\text{i}}\rangle$ of the initial Hamiltonian $H^{\text{i}}$ is time-evolved under the final Hamiltonian $H^{\text{f}}$.
Here two different types of DQPTs can occur: topological DQPTs, whose occurrence is intimately related to the topology of $H^{\text{i}}$ and $H^{\text{f}}$; and accidental DQPTs, which are to the contrary.
Specifically, for quench dynamics of one-dimensional topological systems, topological DQPTs necessarily exist when ground states of $H^\text{i}$ and $H^\text{f}$ belong with distinct topological phases~\cite{Dora15,Balatsky,dutta17}.
These topological DQPTs provide a crucial link between static topological phases and emergent topological phenomena in quench dynamics, and represent an exemplary case where the relation between topology and dynamics can be investigated. In two dimensions, topological DQPTs recently observed in cold atomic gases are associated with dynamic vortices~\cite{,Weitenberg17}, which have been viewed the effective DTOP. In one dimension, on the other hand, relations between DQPTs and DTOPs have yet to be experimentally investigated.

In this work, we report the experimental simulation of topological DQPTs using discrete-time quantum walks (QWs) of single photons in one dimension. We map single-photon QW dynamics to quenches between Floquet topological phases (FTPs)~\cite{CSPRA}, and
probe inner products of the initial and time-evolved states via inference-based measurements.
We then investigate DQPTs by constructing quantities such as the rate function and DTOPs from our measurements.
An advantage of photonic QW dynamics lies in the relative ease of introducing decoherence and loss, which further allows us to experimentally investigate DQPTs for mixed states and in non-unitary quench processes. Consistent with theoretical predictions~\cite{mixeddtop,wyssh}, we find DQPTs persist in unitary quench dynamics of mixed states, as well as in parity-time ($\mathcal{PT}$)-symmetric non-unitary quench dynamics in the $\mathcal{PT}$-symmetry-unbroken regime. In both cases, however, the behaviors of DTOPs are quite different.

\begin{figure*}
\includegraphics[width=\textwidth]{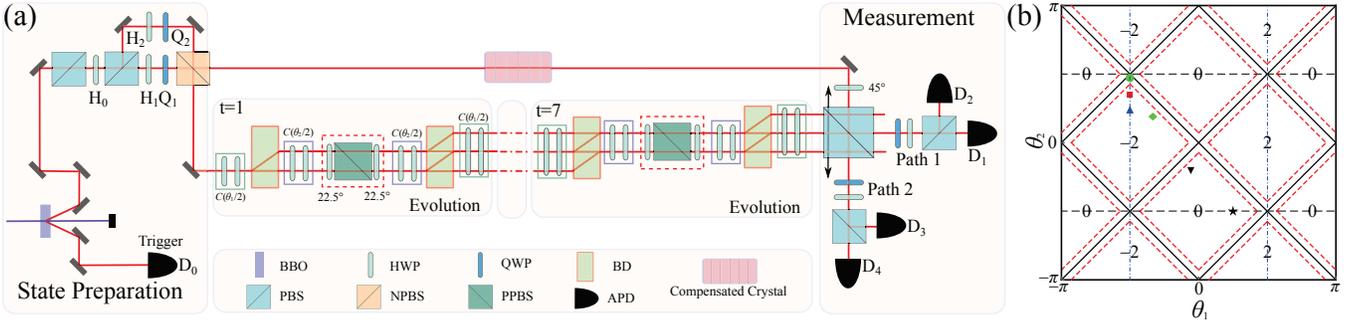}
\caption{(a) Experimental setup for the simulation of DQPTs using QWs. Pairs of single photons are generated via type-I spontaneous parametric down conversion (SPDC) using a non-linear $\beta$-Barium-Borate (BBO) crystal. One photon serves as a trigger and the other signal photon is prepared in an arbitrary linear polarization state using polarizing beam splitters (PBSs), wave plates (WPs) with certain setting angles and a non-polarizing beam splitter (NPBS). Coin rotations and conditional translations are realized by two half-wave plates (HWPs) and a beam displacer (BD), respectively. For non-unitary QWs, a sandwich-type HWP-PPBS-HWP setup is inserted to introduce the partial measurement, where PPBS is an abbreviation for partially polarizing beam splitters. Avalanche photodiodes (APDs) detect the signal and heralding photons.
(b) Phase diagram for QWs governed by Floquet operators $U$ and $\tilde{U}$, labeled by the winding number $\nu$ as a function of coin parameters $(\theta_1,\theta_2)$. Note that topological phase boundaries and winding numbers for unitary QW dynamics governed by $U$ and non-unitary dynamics governed by $\tilde{U}$ are the same. Dashed red lines represent boundaries between $\mathcal{PT}$-symmetry-unbroken and
broken regimes for $\tilde{U}$, with $\mathcal{PT}$-symmetry-broken regimes lying inbetween the red lines near topological phase boundaries, which are represented by solid black lines.
Black star represents coin parameters of the initial Floquet operator $U^\text{i}$ or $\tilde{U}^\text{i}$, other symbols indicate coin parameters of final Floquet operators in different cases.}
\label{fig:setup}
\end{figure*}


{\it Simulating quench dynamics between FTPs:---}
We study DQPTs in quench dynamics using discrete-time QWs on a one-dimensional homogeneous lattice $L$ ($L\in \mathbb{Z}$), where we
use polarization states of single photons $\{\ket{H},\ket{V}\}$ to represent coin states and spatial modes to encode walker states. As illustrated in Fig.~\ref{fig:setup}(a), the main component of our setup is a cascaded interferometric network. The resulting QW dynamics is governed by the Floquet operator
\begin{align}
U=C(\theta_1/2)SC(\theta_2)SC(\theta_1/2).
\label{eqn:U}
\end{align}
Here, the coin operator $C(\theta)$ rotates the single-photon polarization by $\theta$ about the $y$-axis. The shift operator $S$ moves the walker in $\ket{H}$ ($\ket{V}$) to the left (right) by one lattice site.

QWs governed by $U$ support non-trivial FTPs, which are characterized by winding numbers~\cite{KB+12,CSPRA,supp}.
Topological properties of $U$ can be understood by considering the effective Hamiltonian $H_{\text{eff}}$ defined through $U=e^{-i H_{\text{eff}}}$, where, for homogeneous QWs, $H_{\text{eff}}(k)=E_k \bm{h}\cdot\bm{\sigma}$ in quasimomentum~$k$ space. Here  $\bm{\sigma}$ is the Pauli vector, $\pm E_k$ are the quasienergies, and $\bm{h}$ marks the direction of the spinor eigenvector at each quasi-momentum $k$. As $H_{\text{eff}}$ satisfies chiral symmetry with $\Gamma H_{\text{eff}}\Gamma=-H_{\text{eff}}$ and $\Gamma=\sigma_x$, winding numbers are defined as the number of times $\bm{h}$ winds around the $x$-axis as $k$ varies through the first Brillouin zone (1BZ).
As illustrated in Fig.~\ref{fig:setup}(b), by varying coin parameters $(\theta_1,\theta_2)$, the system can change between FTPs with distinct winding numbers.

For a typical QW process, the photon is initialized in a local state $|\psi^\text{i}\rangle$ at $x=0$, and is subject to repeated operations of $U$, such that at the $t$-th step, the photon is in the state $|\psi(t)\rangle=U^t |\psi^\text{i}\rangle=e^{-iH_{\text{eff}}t} |\psi^\text{i}\rangle$.
Importantly, if we choose $|\psi^{\text{i}}\rangle$ to be an eigenstate of $U^\text{i}=e^{-i H^{\text{i}}_{\text{eff}}}$, the resulting QW dynamics realize stroboscopic simulation of quenches between FTPs associated with $H^{\text{i}}_{\text{eff}}$ and $H_{\text{eff}}$, respectively. For the discrete-time QW protocol considered here, Floquet operators $U^\text{i}$ having localized eigenstates exist, whose coin parameters are on the horizontal black dashed lines in Fig.~\ref{fig:setup}(b). The corresponding $H^{\text{i}}_{\text{eff}}$ are topologically trivial with $\nu^{\text{i}}=0$. For contrast, in the following, we label the Floquet operator actually driving the QW as $U^\text{f}$, with $\nu^{\text{f}}$ the corresponding winding number.

{\it Initialization and detection:---}
Experimentally, We initialize the walker photon at $x=0$, with its coin state
given by the density matrix $\rho_0=p\ket{\psi^\text{i}_-}\bra{\psi^\text{i}_-}+(1-p)\ket{\psi^\text{i}_+}\bra{\psi^\text{i}_+}$, where $\ket{\psi^\text{i}_\pm}=(\ket{H}\mp i\ket{V})/\sqrt{2}$. The initial state is therefore a pure state when $p=\{0,1\}$, and a mixed state otherwise. Importantly, $|x=0\rangle\otimes\ket{\psi^\text{i}_{\pm}}$ are eigenstates of $U^{\text{i}}$ with the coin parameters $(\theta^\text{i}_1=\pi/4,\theta^\text{i}_2=-\pi/2)$.
We then implement QWs governed by $U^{\text{f}}$ with coin parameters $(\theta^{\text{f}}_1,\theta_2^{\text{f}})$. To reduce experimental error, we choose $(\theta^{\text{f}}_1,\theta_2^{\text{f}})$ on blue dash-dotted lines in Fig.~\ref{fig:setup}(b), as the spatial spread of the resulting QW dynamics is small.


\begin{figure}
\includegraphics[width=0.5\textwidth]{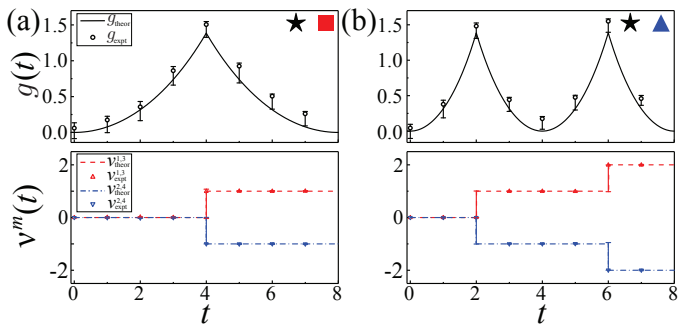}
\caption{Rate function (upper layer) and $\nu^m(t)$ (lower layer) of seven-step unitary QWs as functions of time steps.
The initial state of the walker-coin system is $\ket{x=0}\otimes\ket{\psi^\text{i}_-}$.
QWs are governed by $U^\text{f}$ with $(\theta^\text{f}_1=-\pi/2,\theta^\text{f}_2=3\pi/8)$ (a) and $U^\text{f}$ with $(\theta^\text{f}_1=-\pi/2,\theta^\text{f}_2=\pi/4)$ (b), respectively. Error bars are derived from simulations where we consider all the systematic inaccuracies of the experiment.}
\label{fig:pure}
\end{figure}

Due to the lattice-translational symmetry, time evolutions in different quasimomentum~$k$-sectors are decoupled and governed by $U^\text{f}_k$, the Fourier component of $U^\text{f}$. We construct the Loschmidt amplitude $G(k,t)$ in each quasimomentum~$k$-sector according to
\begin{align}
G(k,t):=\text{Tr}\left[\rho_0\left(U^\text{f}_k\right)^t\right]=\sum_{x}\text{e}^{-ikx}\bar{P}(x,t),
\end{align}
where $\bar{P}(p,x,t)=p\langle\psi^\text{i}_-|\psi_-(x,t)\rangle+(1-p)\langle\psi^\text{i}_+|\psi_+(x,t)\rangle$, and
$|\psi_\pm(x,t)\rangle=\sum_k e^{ikx}\left(U^\text{f}_k\right)^t|\psi^\text{i}_\pm\rangle$.
Experimentally, $\bar{P}(p,x,t)$ is measured by performing interference-based measurements at the $t$-th step~\cite{supp,NY17}.
More specifically, after the photons pass through the QW interferometric network, we project the polarization state of the photons at each position $x$ onto the initial polarization state of the photons at $x=0$, and perform coincidence measurements on the walker photons and the trigger photons successively up to $t$ by single-photon avalanche photodiodes (APDs).

We then construct the rate function according to $g(t)=-\sum_{k\in \text{1BZ}}\ln |G(k,t)|^2$, where we have used $G(t)=\prod_{k\in \text{1BZ}} G(k,t)$. Hence by construction, $g(t)$ corresponds to the rate function of a quench starting from a half-filled Floquet band, where the initial state is a direct product of single-particle density matrices $\rho_0$ in different $k$-sectors of 1BZ. This is in contrast to the case of single-photon QW dynamics, where the initial state is a superposition of coin states in different $k$-sectors.
We therefore emphasize that whereas DQPTs do not actually occur in QWs of single photons, we can simulate DQPTs in quench dynamics of topological systems using the setup.

From the measured $G(k,t)$, we further calculate DTOPs characterizing DQPTs. In one dimension, DTOPs are defined as~\cite{BH16}
\begin{align}
\nu^m(t)=\frac{1}{2\pi}\int_{k_m}^{k_{m+1}}\frac{\partial\phi_{k}^{G}(t)}{\partial k}\text{d}k,
\end{align}
where the Pancharatnam geometric phase (PGP) $\phi_{k}^{G}(t)=\phi_{k}(t)-\phi_{k}^{\text{dyn}}(t)$. Here $\phi_{k}(t)$ is defined through $G(k,t)=|G(k,t)|e^{i \phi_{k}(t)}$, and $\phi_{k}^{\text{dyn}}(t)$ is the dynamic phase. $k_m$ ($m=1,2,...$) are fixed points of the dynamics, where the corresponding density matrices do not evolve in time and PGP vanishes at all times. $\nu^m(t)$ therefore characterizes the $S^1\rightarrow S^1$ mapping from the momentum submanifold between $k_m$ and $k_{m+1}$ to $e^{i\phi_k^{G}(t)}$. The DTOP is quantized and can only change value at DQPTs, where $G(k_c,t_c)=0$ and $\phi_{k_c}^{G}(t_c)$ becomes ill-defined at critical $k_c$ and $t_c$. While fixed points necessarily exist when $U^\text{i}$ and $U^\text{f}$ have different winding numbers, $k_c$ exists between adjacent fixed points and leads to DQPTs at $t_c=(2n-1)t_0$ ($n\in \mathbb{N}$), where the critical time scale $t_0=\pi/(2E^\text{f}_{k_c})$ and $\pm E^\text{f}_k$ is the quasienergy of $U^\text{f}_k$. As a result, $\nu^m(t)$ exhibits abrupt jumps at the DQPTs associated with $k_c\in (k_m,k_{m+1})$.


\begin{figure}
\includegraphics[width=0.5\textwidth]{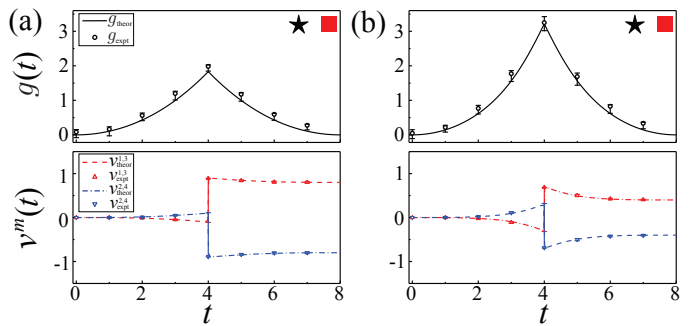}
\caption{Rate function (upper layer) and $\nu^m(t)$ (lower layer) of a seven-step unitary QW. The walker starts at $x=0$ and the QW is governed by the final Floquet operator $U^\text{f}$ with $(\theta^\text{f}_1=-\pi/2,\theta^\text{f}_2=3\pi/8)$. The initial coin state is a mixed state with $p=0.9$ (a) and $p=0.7$ (b), respectively. Error bars are derived from simulations where we consider all the systematic inaccuracies of the experiment.}
\label{fig:mixed}
\end{figure}

\begin{figure}
\includegraphics[width=0.5\textwidth]{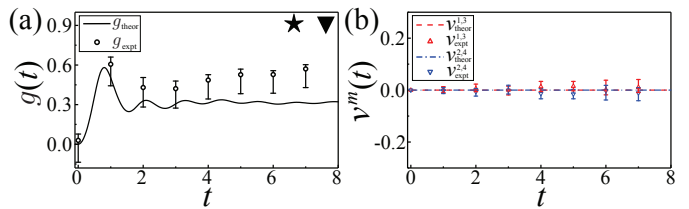}
\caption{(a) Rate function and (b) $\nu^m(t)$ of a seven-step unitary QW. The QW is governed by the final Floquet operator $U^\text{f}$ with $(\theta^\text{f}_1=-\pi/16,\theta^\text{f}_2=-3\pi/16)$, which is in the same FTP as the initial state. Error bars are derived from simulations where we consider all the systematic inaccuracies of the experiment.}
\label{fig:trivial}
\end{figure}

\begin{figure*}
\includegraphics[width=0.75\textwidth]{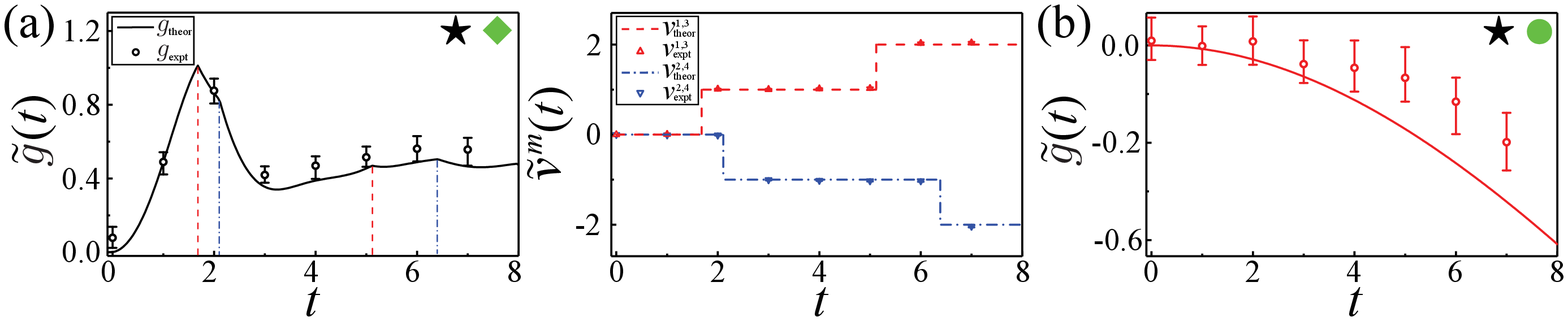}
\caption{(a) Rate function and $\tilde{\nu}^m(t)$ of a seven-step non-unitary QW with a loss parameter $l=0.36$. The initial state of walker-coin system is $\ket{0}\otimes\ket{\psi_-}$. The QW is governed by the non-unitary Floquet operator $\tilde{U}^\text{f}$ with $(\theta^\text{f}_1=-\pi/3,\theta^\text{f}_2=\pi/5)$. (b) Rate function of the QW governed by $\tilde{U}^\text{f}$ with $\left[\theta^\text{f}_1=-\pi/2,\theta^\text{f}_2=(\pi-\xi)/2\right]$, where $\xi=\arccos(1/\alpha)$ and $\alpha=(1+\sqrt{1-l})/(2\sqrt[4]{1-l})$. The two critical time scales are $t_0=\{1.7183,2.1482\}$, which give rise to non-analyticities in the rate function as indicated by vertical dashed lines in (a). Theoretically calculated fixed points are located at $k_{1,2,3,4}=\{-1.0094\pi,-0.4470\pi,-0.0094\pi,0.5530\pi\}$, and the critical momenta $k_c=\{-0.7888\pi,-0.1534\pi,0.2112\pi,0.8466\pi\}$. Experimental errors are due to photon-counting statistics.}
\label{fig:nonunitary}
\end{figure*}

{\it DQPT in unitary dynamics:---}
We first study DQPTs for pure states in unitary dynamics. We initialize photons in the coin state $\ket{\psi^\text{i}_-}$ at $x=0$. The photons are then subject to unitary time evolutions governed by the Floquet operator $U^\text{f}$ with $(\theta^\text{f}_1=-\pi/2,\theta^\text{f}_2=3\pi/8)$. This corresponds to a quench between FTPs with $\nu^{\text{i}}=0$ and $\nu^{\text{f}}=-2$. Here, the fixed points $k_{1,2,3,4}=\{-\pi,-\pi/2,0,\pi/2\}$, and $k_c=\{\pm \pi/4,\pm 3\pi/4\}$. Note $U^\text{f}_k$ has a discrete symmetry $U^\text{f}_k=U^\text{f}_{k+\pi}$ in addition to the time-reversal symmetry. Under these symmetries, $E^\text{f}_{k_c}$ are degenerate and there is only one critical time scale $t_0=4$. In Fig.~\ref{fig:pure}(a), we show the rate function, which becomes non-analytic at the first critical time $t_c=t_0$.
Whereas it is difficult to directly identify non-analyticities of $g(t)$ in discrete-time dynamics, DQPTs are unambiguously revealed by jumps in the quantized DTOP across $t_c$. The abrupt jump is confirmed by particularly large error bars in the measured $\nu^m(t)$ at the critical time~\cite{supp}. Further, due to the symmetry of $U^\text{f}_k$, we have $\nu^{1,3}(t)=-\nu^{2,4}(t)$, where $\nu^{4}(t)$ is integrated in the range $(\pi/2,\pi)$.

We then fix the initial coin state and change the final Floquet operator to $U^\text{f}$ with $(\theta^\text{f}_1=-\pi/2,\theta^\text{f}_2=\pi/4)$. As shown in Fig.~\ref{fig:pure}(b), the critical time scale changes to $t_0=2$, and the rate function becomes non-analytic at odd multiples of $t_0$.
The quantized DTOPs also feature abrupt jumps at critical times.
As locations of $k_m$ and $k_c$ are the same as those of the previous case, there is only one critical time scale as well.

In the second case study, we initialize photons at $x=0$ and in a mixed coin state characterized by $\rho_0$ with $p=0.7$ and $p=0.9$, respectively. The QW is governed by $U^\text{f}$ with $(\theta^\text{f}_1=-\pi/2,\theta^\text{f}_2=3\pi/8)$. Whereas the resulting QW dynamics still correspond to quenches between FTPs with $\nu^{\text{i}}=0$ and $\nu^{\text{f}}=-2$, coin states of time-evolved states remain mixed. We show the resulting rate functions and DTOPs in Fig.~\ref{fig:mixed}.
Whereas the occurrence of DQPTs are still signaled by non-analyticities in the rate functions, DTOPs are no longer quantized. This is because PGPs do not vanish at $k_m$, such that $e^{i\phi^G_k}$ no longer forms a closed $S^1$ manifold between $k_m$ and $k_{m+1}$. Consequently, $\nu^m(t)$ is no longer the winding number characterizing such a map. These results are consistent with previous theoretical studies~\cite{mixeddtop}. Note locations of $k_m$ and $k_c$ are the same as in the previous cases.

For comparison, we also study the case where the quench dynamics is between FTPs with $\nu^{\text{i}}=0$ and $\nu^{\text{f}}=0$.
For this purpose, we choose $U^\text{f}$ with $(\theta^\text{f}_1=-\pi/16,\theta^\text{f}_2=-3\pi/16)$. As shown in
Fig.~\ref{fig:trivial}, the rate function is smooth in time and $\nu^m(t)$ remains zero, indicating the absence of DQPTs. Here $k_m=\{0,\pm\pi/2,\pi\}$.

{\it DQPT in $\mathcal{PT}$-symmetric non-unitary dynamics:---}
The ease of introducing loss in photonics further allows us to explore DQPTs in non-unitary dynamics~\cite{wyssh,ZWWG17}.
We enforce non-unitary dynamics by performing a partial measurement $M_e=\one_w\otimes\sqrt{l}\ket{-}\bra{-}$ in the basis $\{\ket{\pm}\}$ at each time step, with $\one_w=\sum_x\ket{x}\bra{x}$, $\ket{\pm}=(|H\rangle\pm |V\rangle)/\sqrt{2}$, and $l$ the loss parameter, which is fixed at $l=0.36$ in our experiment. The non-unitary QW is then governed by
\begin{align}
\tilde{U}=\gamma C(\theta_1/2)SC(\theta_2/2)MC(\theta_2/2)SC(\theta_1/2),
\end{align}
where $M=\one_\text{w}\otimes\left(\ket{+}\bra{+}+\sqrt{1-l}\ket{-}\bra{-}\right)$, and $\gamma=(1-l)^{-1/4}$.

Topological properties of $\tilde{U}$ are characterized by winding numbers defined through the global Berry phase~\cite{GW88,LH13,Lieu18}. The resulting topological phase diagram is the same as that of $U$~\cite{supp}.
Crucially, $\tilde{U}$ also possess $\mathcal{PT}$ symmetry, therefore its quasienergy spectra can be entirely real in the $\mathcal{PT}$-symmetry-unbroken regime~\cite{BB98,BBJ02,B07}. In the regime with spontaneously broken $\mathcal{PT}$ symmetry, on the other hand, quasienergies of $\tilde{U}$ can be complex. The boundary between regimes with unbroken and broken $\mathcal{PT}$ symmetry is plotted in Fig.~\ref{fig:setup}(b) as red dashed lines, with $\mathcal{PT}$-symmetry-broken regimes surrounding topological phase boundaries.
It can be shown that DQPTs necessarily occur for quench processes between distinct FTPs in the $\mathcal{PT}$-symmetry-unbroken regime~\cite{wyssh,supp}.


Similar to the unitary case, we initialize photons in the state $|x=0\rangle\otimes\ket{\psi^\text{i}_-}$, with the corresponding $\tilde{U}^\text{i}$ in the $\mathcal{PT}$-symmetry-unbroken regime with $\nu^\text{i}=0$.
The walker is evolved under the final non-unitary Floquet operator $\tilde{U}^\text{f}$ with $(\theta^\text{f}_1=-\pi/3, \theta^\text{f}_2=\pi/5)$, which is in the $\mathcal{PT}$-symmetry-unbroken regime with $\nu^\text{f}=-2$. The Loschmidt amplitude, the rate function $\tilde{g}(t)$, and the DTOP $\tilde{\nu}^m(t)$ can be constructed similar to the unitary case~\cite{supp}, albeit QW dynamics is now non-unitary. As illustrated in Fig.~\ref{fig:nonunitary}(a), non-analyticities in the rate function have two distinct time scales, which correspond to two different DTOPs [see Fig.~\ref{fig:nonunitary}(b)], both quantized and demonstrating abrupt jumps at odd multiples of the corresponding critical time scale.

The emergence of two critical time scales is intimately connected with the breaking of time-reversal symmetry of the non-unitary dynamics~\cite{wyssh}.
In this case, whereas fixed points still exist when $U^\text{i}$ and $U^\text{f}$ are in the $\mathcal{PT}$-symmetry-unbroken regime and feature distinct FTPs, they are no longer located at high-symmetry points and have to be solved numerically. As a consequence, $\nu^{1,3}(t)$ and $\nu^{2,4}(t)$ feature jumps at different critical times, giving rise to two critical time scales.


Finally, we study the case when the final non-unitary Floquet operator is in the $\mathcal{PT}$-symmetry-broken regime. The resulting rate function is shown in Fig.~\ref{fig:nonunitary}(b), where no DQPTs can be identified. As fixed points are also absent in the dynamics, DTOPs cannot be defined in this case~\cite{wyssh}.

{\it Conclusions:---}
We have demonstrated photonic QWs as an ideal platform for the simulation of DQPTs in quench dynamics of topological systems. By constructing key quantities such as the rate function and DTOPs from interference measurements, we experimentally investigate the relation between DQPTs and DTOPs in both unitary and non-unitary quench dynamics.
Our experiment opens up the avenue of investigating dynamic topological phenomena using QW dynamics, and paves the way for a more systematic study of DQPTs in novel situations such as engineered non-unitary dynamics or in higher dimensions.

\acknowledgments
\textit{Acknowledgement:--}
This work has been supported by the Natural Science Foundation of China (Grant Nos. 11474049, 11674056, and 11522545) and the Natural Science Foundation of Jiangsu Province (Grant No. BK20160024). WY acknowledges support from the National Key R\&D Program (Grant Nos. 2016YFA0301700,2017YFA0304100). KW and XQ contributed equally to this work.

\clearpage
\begin{widetext}
\appendix

\renewcommand{\thesection}{\Alph{section}}
\renewcommand{\thefigure}{S\arabic{figure}}
\renewcommand{\thetable}{S\Roman{table}}
\setcounter{figure}{0}
\renewcommand{\theequation}{S\arabic{equation}}
\setcounter{equation}{0}

\section{Supplemental Materials}

\section{Experimental implementation}
The main component of our setup is the cascaded interferometric network as shown in Fig.~\ref{fig:setup}(a) of the main text. For the single-photon source, we use $401.8$nm continuous-wave diode laser to pump an optically nonlinear $\beta$-Barium-Borate (BBO) crystal. The polarization-degenerate photon pairs are generated by the non-collinear type-I spontaneous parametric down conversion (SPDC). The trigger photon heralds the presence of a signal photon.

We implement the coin operator $C(\theta)=\sum_x\ket{x}\bra{x}\otimes e^{-i\theta\sigma_y}$ via two half-wave plates (HWPs). The angle of the first HWP is set to $0$ and that of the second is $\theta$. The shift operator $S$ via a beam displacer (BD) whose optical axis is cut so that the photons in $\ket{V}$
are directly transmitted and those in $\ket{H}$ undergo a lateral displacement into a neighboring mode~\cite{pxprl}.

To realize a non-unitary quantum walk (QW), the non-unitary dynamics is enforced by performing a partial measurement $M_e=\one_w\otimes\sqrt{l}\ket{-}\bra{-}$ in the basis $\{\ket{\pm}=(\ket{H}\pm\ket{V})/\sqrt{2}\}$ at each step, with $\one_w=\sum_x\ket{x}\bra{x}$ and $l$ the loss parameter and is fixed to $0.36$ in our experiment. The partial measurement operator $M_e$ is realized by a sandwich-type setup involving two HWPs at $22.5^\circ$ and a partially polarizing beam splitter (PPBS). At each step, after applying the partial measurement, photons in the state $\ket{-}$ are reflected by the PPBS with probability $l$ and the rest photons continue propagating in the quantum-walk dynamics. By choosing proper coin parameters $(\theta^\text{f}_1,\theta^\text{f}_2)$, the non-unitary QW can possess parity-time ($\mathcal{PT}$) symmetry. By changing coin parameters, the non-unitary QW can be either in the $\mathcal{PT}$-symmetry-unbroken or -broken regimes.

In our experiment, the initial coin state is chosen as either a pure state such as $\ket{\psi^\text{i}_-}$ or a mixed state such as $\rho_0=p\ket{\psi^\text{i}_-}\bra{\psi^\text{i}_-}+(1-p)\ket{\psi^\text{i}_+}\bra{\psi^\text{i}_+}$, where $\ket{\psi^\text{i}_\pm}=(\ket{H}\mp i\ket{V})/\sqrt{2}$. As $\ket{\psi^\text{i}_-}$ is a special case of $\rho_0$ with $p=1$, we only show how to prepare the signal photon in a mixed state. After photons passing the first polarizing beam splitter (PBS), only the horizontally polarized photons are injected into the remaining optical setup. The first HWP (H$_0$) is set to $\theta=\arccos (\sqrt{p})/2$. Hence, we are able to prepare an arbitrary mixed state by controlling the setting angle of H$_0$.
For the pure initial state $\ket{\psi^\text{i}_-}$, the setting angle of H$_0$ is taken to be $0$. The second PBS splits the photons polarization-wise into two spatial paths, where photons are prepared in either $\ket{\psi^\text{i}_+}$ or $\ket{\psi^\text{i}_-}$ with HWPs (H$_1$ or H$_2$) and quarter-wave plates (QWPs, Q$_1$ or Q$_2$). The optical-path difference between two paths is longer than the coherence length of photons. Finally, a $50:50$ non-polarizing beam splitter (NPBS) combines photons from two paths and then re-splits them into two paths. Photons in one path are injected into the cascaded interferometric network which implements QWs. Photons in the other path are in the initial state, and are used for measurement after the phases caused by the optical difference between the two paths are compensated.

\section{Interference-based measurements of inner products}
In quasimomentum space,
\begin{align}
G(k,t)&=\text{Tr}\left[\rho_0\left(U^\text{f}_k\right)^t\right]=p\langle\psi^\text{i}_-|\psi_-(t)\rangle+(1-p)\langle\psi^\text{i}_+|\psi_+(t)\rangle\nonumber\\
&=\sum_{x}\text{e}^{-ikx}\bar{P}(x,t),\nonumber
\end{align}
where $\bar{P}(p,x,t)=p\langle\psi^\text{i}_-|\psi_-(x,t)\rangle+(1-p)\langle\psi^\text{i}_+|\psi_+(x,t)\rangle$, $|\psi_\pm(t)\rangle=\left(U^\text{f}_k\right)^t|\psi^\text{i}_\pm\rangle$, and $|\psi_\pm(x,t)\rangle=\sum_k e^{ikx}|\psi_{\pm}(t)\rangle$ is the coin state on site $x$ at time $t$ when the initial state is $|x=0\rangle\otimes |\psi^\text{i}_\pm\rangle$. The trace is taken with respect to the coin state in the $k$-sector.

In this section, we discuss how to experimentally measure $\bar{P}(p,x,t)$ through interference-based measurements.
Without loss of generality, we write
\begin{align}
&\ket{\psi^\text{i}_-}=(\alpha,\beta)^\text{T}, \ket{\psi^\text{i}_+}=(a,b)^\text{T},\\
&\ket{\psi_-(x,t)}=(\alpha',\beta')^\text{T}, \ket{\psi_+(x,t)}=(a',b')^\text{T}. \nonumber
\end{align}
The polarization state of the photons after the $t$-th steps at $x$ is given by the density matrix
\begin{equation}
\rho_t(x)=p\ket{\psi_-(x,t)}\bra{\psi_-(x,t)}+(1-p)\ket{\psi_+(x,t)}\bra{\psi_+(x,t)}.
\end{equation}

At the measurement stage, the first PBS combines the photons in the states $\rho_0$ and $\rho_t(x)$ and re-splits them into paths 1 and 2 depending on their polarizations. The polarization states of the photons in paths 1 and 2 are
\begin{align}
&\rho_{P1}=p\begin{pmatrix}
     \alpha'^2 & \alpha'\alpha^* \\
     \alpha\alpha'^* & \alpha^2  \\
     \end{pmatrix}+(1-p)\begin{pmatrix}
     a'^2 & a'a^* \\
     aa'^* & a^2  \\
     \end{pmatrix},\\
&\rho_{P2}=p\begin{pmatrix}
     \beta^2 & \beta\beta'^* \\
     \beta'\beta^* & \beta'^2  \\
     \end{pmatrix}+(1-p)\begin{pmatrix}
     b^2 & bb'^* \\
     b'b^* & b'^2  \\
     \end{pmatrix},\nonumber
\end{align}
respectively.

Using a QWP, a HWP and a PBS, we apply the projector $\ket{\psi^\text{i}_+}\bra{\psi^\text{i}_+}$ on photons in path 1. The probability
\begin{align}
P_{11}=\frac{1}{2}\left[p\alpha'^2+(1-p)a'^2+p\alpha^2+(1-p)a^2\right]+\text{Im}\left[p\alpha'\alpha^*+(1-p)a'a^*\right]
\end{align}
can be read out from the coincidence between the detectors D$_1$ and D$_0$ (to record the triggering photons). Similarly, the probability $P'_{11}$ is read out from the coincidence between D$_2$ and D$_0$. Thus, the probability of photons measured in path 1 is
\begin{equation}
P_1=P_{11}+P'_{11}=p\alpha'^2+(1-p)a'^2+p\alpha^2+(1-p)a^2.
\end{equation}
By changing the setting angles of the QWP and HWP, we apply the other projector $\ket{+}\bra{+}$ on photons in path 1. The resulting probability is
\begin{align}
P_{12}=\frac{1}{2}\left[p\alpha'^2+(1-p)a'^2+p\alpha^2+(1-p)a^2\right]+\text{Re}\left[p\alpha'\alpha^*+(1-p)a'a^*\right],
\end{align}
which can be read out from the coincidence between the detectors D$_1$ and D$_0$.

Similarly, we obtain the probability of photons in path 2 by measuring in the basis state $\ket{\psi^\text{i}_+}$ and $\ket{+}$, i.e., $P_{21}$ and $P_{22}$ respectively. These can be read out from the coincidence between D$_4$ and D$_0$. With the probability $P'_{21}$ read out from the coincidence between D$_3$ and D$_0$, we have $P_2=P_{21}+P'_{21}$. Hence, we can calculate $\bar{P}(p,x,t)$ from the directly measured probabilities as
\begin{align}
\bar{P}(p,x,t)=i\left(P_{11}-\frac{P_1}{2}-P_{21}+\frac{P_2}{2}\right)+\left(P_{12}-\frac{P_1}{2}+P_{22}-\frac{P_{2}}{2}\right).
\end{align}

From $\bar{P}(p,x,t)$, we then construct the rate function and DTOPs as outlined in the main text.

\section{Error analysis}

We have four sources of systematic errors in our experimental setup: imperfections of optical elements, difference in photon loss in different optical paths, statistical noise, and decoherence in the interference-based measurements.

First, the accuracy of angles of WPs is about $\pm0.1^\circ$. Second, losses in different optical paths split by the NPBS are estimated in an independent measurement with an accuracy of $\pm2\%$. Third, errors due to photon-counting statistics are scaled by the square root of the number of click events. Here, total coincidence counts are about $40,000$ over a collection time of $20$s. For unitary QWs, errors due to Poissonian statistics are a minor contribution compared to other systematic deviations. However, for non-unitary QWs, as non-unitarity is introduced via loss of photons, total coincidence counts decrease, and errors due to photon-counting statistics become significant compared to other errors. Thus in Fig.~\ref{fig:nonunitary}, error bars are derived only from photon-counting statistics. Fourth, in our experiment, the major source of decoherence is dephasing, caused by the optical-path difference between paths split by NPBS. We assume that the dephasing rate $\eta=0.97$ is a constant to simplify estimation. Under dephasing, the time-evolved coin state is not pure anymore. We estimate the density matrix of the coin state as $\varepsilon(\rho_c)=\eta\rho_c+(1-\eta)\sigma_z\rho_c\sigma_z$, where $\sigma_z$ is the standard Pauli operator.

Taking into account all possible combinations of systematic errors listed above, we use experimentally measured values and estimated error ranges as input for Monte Carlo simulations. We then use analytic results as reference, and take the largest positive (negative) deviation of numerical results from the reference as the negative (positive) error bar. Due to decoherence, the error bars are typically asymmetric. We also note that at critical times where $\nu^m(t)$ undergo abrupt jumps, the error bars also become large and asymmetric, as the system dynamics become sensitive to small perturbations at these times.

\section{Topological invariants for QW dynamics}
In this section, we discuss the calculation of winding numbers for $U$ and $\tilde{U}$ defined in the main text.
First, we consider the unitary QW dynamics governed by $U$.
Taking the Fourier transform, we derive the Floquet operator $U_k$ in each $k$-sector
\begin{equation}
\begin{split}
U_k = &~ n_0\sigma_0-n_1\sigma_1-in_2\sigma_2-in_3\sigma_3,\\
n_0 =&~ \cos(2k)\cos\theta_1\cos\theta_2-\sin\theta_1\sin\theta_2,\\
n_1 =&~  0,\\
n_2 =&~ \cos(2k)\cos\theta_2\sin\theta_1+\cos\theta_1\sin\theta_2,\\
n_3 =&~ -\sin(2k)\cos\theta_2.
\end{split}\label{eqn:unitaryn}
\end{equation}
Topological properties of $U$ are characterized by the winding number, defined as
\begin{align}
\nu=-\frac{1}{2\pi}\int_{-\pi}^{\pi} \left(\bm{n}\times \frac{\text{d}\bm{n}}{\text{d}k}\right)_1.
\end{align}
Winding number calculated in this way is equivalent to that from the Bloch vector of the effective Hamiltonian.

In non-unitary QW dynamics governed by $\tilde{U}$, topologically inequivalent phases can be distinguished by the winding number $\nu=\varphi_\text{B}/2\pi$. Here, the global Berry phase $\varphi_{\text{B}}$ is
\begin{align}
\varphi_B=-i\sum_{\mu=\pm}\int_{-\pi}^{\pi} \text{d}k\frac{\langle\chi_{\mu}|\frac{\partial}{\partial k}|\psi_{\mu}\rangle}{\langle\chi_{\mu}|\psi_{\mu}\rangle},
\end{align}
where the right (left) eigenvector is defined as $U_k|\psi_\mu\rangle=\lambda_\mu|\psi_\mu\rangle$ ($U_k^{\dag}|\chi_{\mu}\rangle=\lambda^*_{\mu}|\chi_{\mu}\rangle$).

Topological invariants defined through the global Berry phase $\varphi_B$ have the advantage that they invoke the formalism of biorthonormal basis and characterize topological properties in both the $\mathcal{PT}$-symmetry-preserving and broken regimes.

\section{Parity-time symmetry}
$\tilde{U}=F(\gamma M)G$, where $\gamma=(1-l)^{-1/4}$, has $\mathcal{PT}$ symmetry, where $\mathcal{PT} \tilde{U}(\mathcal{PT})^{-1}=\tilde{U}^{-1}$ with the symmetry operator $\mathcal{PT}=\sum_x|-x\rangle\langle x|\otimes\sigma_z\mathcal{K}$ and $\mathcal{K}$ the complex conjugation. In momentum space, $\tilde{U}$ can be resolved into the following expression
\begin{equation}
\begin{split}
\tilde{U} =&~ d_0\sigma_0-id_1\sigma_1-id_2\sigma_2-id_3\sigma_3,\\
d_0 =&~ \alpha(\cos(2k)\cos\theta_1\cos\theta_2-\sin\theta_1\sin\theta_2),\\
d_1 =&~ i\beta,\\
d_2 =&~ \alpha(\cos(2k)\cos\theta_2\sin\theta_1+\cos\theta_1\sin\theta_2),\\
d_3 =&~ -\alpha\sin(2k)\cos\theta_2,
\end{split}
\label{eqn:U_nonunitary}
\end{equation}
where $\alpha=\frac{\gamma}{2}(1+\sqrt{1-l}), \beta=\frac{\gamma}{2}(1-\sqrt{1-l})$. The eigenvalues of $\tilde{U}$ is $\tilde{\lambda}_{\pm}=d_0\mp i\sqrt{1-d_0^2}$, and the quasienergy $i\ln(\tilde{\lambda}_{\pm})$. When $d_0^2<1$ for all $k$, the quasienergy spectra is real, and the system is in the $\mathcal{PT}$-symmetry-unbroken regime. If $d_0^2\geq1$ for some $k$, the $\mathcal{PT}$ symmetry is spontaneously broken and the quasienergy can be complex.

\section{DQPT for pure states}
In this section, we discuss the simulation of topological DQPTs for pure states using QW dynamics.
Due to the lattice translational symmetry, time evolutions in different $k$-sectors are decoupled. In each $k$-sector, we consider the QW dynamics initialized in the coin state $|\psi^\text{i}_-\rangle$, and governed by the Floquet operator $U^\text{f}_k$. We require that $|\psi^\text{i}_-\rangle$ should be an eigenstate of the Floquet operator $U^\text{i}_k$, and write $U_k|\psi^\text{f}_\pm\rangle=e^{\mp iE^\text{f}_k}|\psi^\text{f}_\pm\rangle$, where $|\psi^\text{f}_\pm\rangle$ are eigenstates of $U^\text{f}_k$ and $\pm E^\text{f}_k$ are the corresponding quasienergy.
With these, the time-evolved state is
\begin{align}
|\psi_-(t)\rangle=~(U^\text{f}_k)^t|\psi^\text{i}_-(k)\rangle=e^{iE^\text{f}_kt}c_-|\psi^\text{f}_-\rangle+e^{-iE^\text{f}_kt}c_+|\psi^\text{f}_+\rangle,
\end{align}
where $c_-=\langle\psi^\text{f}_-|\psi^\text{i}_-\rangle$ and $c_+=\langle\psi^\text{f}_+|\psi^\text{i}_-\rangle$.

The Loschmidt amplitude is then
\begin{align}
G(k,t)=\langle\psi^\text{i}_-|\psi_-(t)\rangle=e^{iE^\text{f}_kt}|c_-|^2+e^{-iE^\text{f}_kt}|c_+|^2:=|G(k,t)|e^{i\phi(k,t)}.
\end{align}
The PGP is defined through $\phi^\text{G}(k,t)=\phi(k,t)-\phi^{\text{dyn}}(k,t)$, where $\phi^{\text{dyn}}(k,t)=E^\text{f}_kt(|c_-|^2-|c_+|^2)$ is the dynamical phase. It is straightforward to show that when $c_+(k_m)=0$ or $c_-(k_m)=0$, PGP vanishes at all times. Further, one can show that at these $k_m$, the corresponding density matrices do not evolve in time. We therefore identify these momenta as fixed points of dynamics.

DQPTs are caused by dynamic Fisher zeros, where the Loschmidt amplitude $G(k_c,t_c)$
vanishes at the time $t_c$. Here $k_c$ is the critical momentum, determined by $|c_{-}(k_c)|^2=|c_+(k_c)|^2$. It follows that DQPTs occur at $t_c=(2n-1)t_0$, where $t_0=\pi/2E^{\text{f}}_{k_c}$ is the critical time scale. Notably, as we will show later in the Supplemental Material, when the system is quenched between different topological phases, fixed points with $c_+=0$ and those with $c_-=0$ always emerge in pairs.
As $|c_+|-|c_-|$ are continuous functions of $k$, there must be at least one critical momentum satisfying $|c_+(k_c)|=|c_-(k_c)|$ inbetween two fixed points of different kinds. DQPTs necessarily exist in this case.

We now consider the more concrete case discussed in the main text, where the initial and final Floquet operators are parameterized by
$(\theta^\text{i}_1=\pi/4,\theta^\text{i}_2=-\pi/2)$ and $(\theta^\text{f}_1=-\pi/2,\theta^\text{f}_2\in(0,\pi/2))$, respectively. In this case, eigenstates of $U^\text{i}$ are spatially localized on the lattice, and $|\psi^\text{i}_-\rangle=(|H\rangle+i|V\rangle)/\sqrt{2}$ for any $k$.
It follows that
$|c_-|^2=\frac{1}{2}(1+\cos2k)$ and $|c_+|^2=\frac{1}{2}(1-\cos2k)$. There are four fixed points at $k_{1,2,3,4}=\{-\pi,-\pi/2,0,\pi/2\}$, and there are four critical momenta at $k_c=\{\pm \pi/4,\pm 3\pi/4\}$. Importantly, due to symmetries of the unitary Floquet operator $U^\text{f}_k$, $E^\text{f}_{k_c}$ are degenerate at different $k_c$. Hence there is only one critical time scale $t_0=\pi/(2E^\text{f}_{k_c})$.

We define the rate function
\begin{equation}
g(t)=-\frac{1}{\pi}\int^{\pi}_{-\pi}\text{d}k\ln |G(k,t)|,\label{eqn:supprate}
\end{equation}
where $N$ is the system size. As illustrated in Fig.~S1, DQPTs occur when $g(t)$ becomes non-analytical at critical times.

DQPTs are characterized by DTOPs, defined in terms of the PGP as
\begin{align}
\nu^m(t)=\frac{1}{2\pi}\int^{k_{m+1}}_{k_m}\text{d}k\frac{\partial\phi^G(k,t)}{\partial k}.
\end{align}
As the PGP $\phi^G(k,t)$ vanishes $k_m$, $\nu^m(t)$ characterizes the $S^1\rightarrow S^1$ mapping from the momentum submanifold between $k_m$ and $k_{m+1}$ to $e^{i\phi_k^{G}(t)}$. $\nu_m(t)$ is therefore quantized, and can only change values at critical times, when $G(k_c,t_c)=0$ and $\phi^G(k_c,t_c)$ becomes ill-defined (see Fig.~S1).

Finally, we emphasize that the rate function constructed in Eq.~(\ref{eqn:supprate}) corresponds to that of a system
initialized in the state $\prod_{k\in \text{1BZ}}|\psi^\text{i}_-\rangle$. We are therefore simulating quench dynamics of a many-body system using single-photon QW dynamics.

\begin{figure}
\includegraphics[width=16cm]{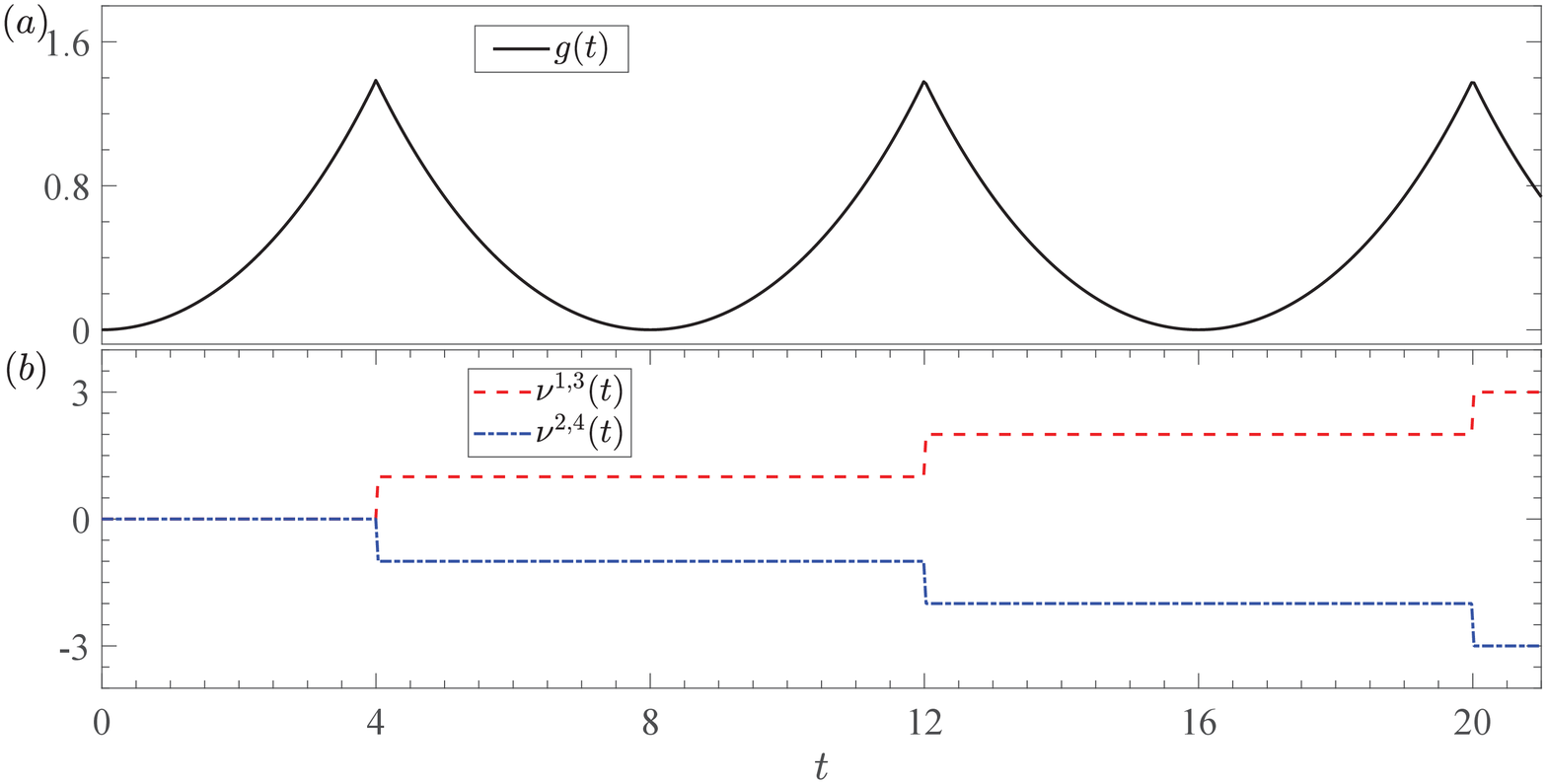}
\caption{Rate function (a) and DTOPs (b) for unitary dynamics of pure states. The coin parameters are the same as those in Fig.~\ref{fig:pure}(a) of the main text.}
\end{figure}

%

\begin{figure}
\includegraphics[width=16cm]{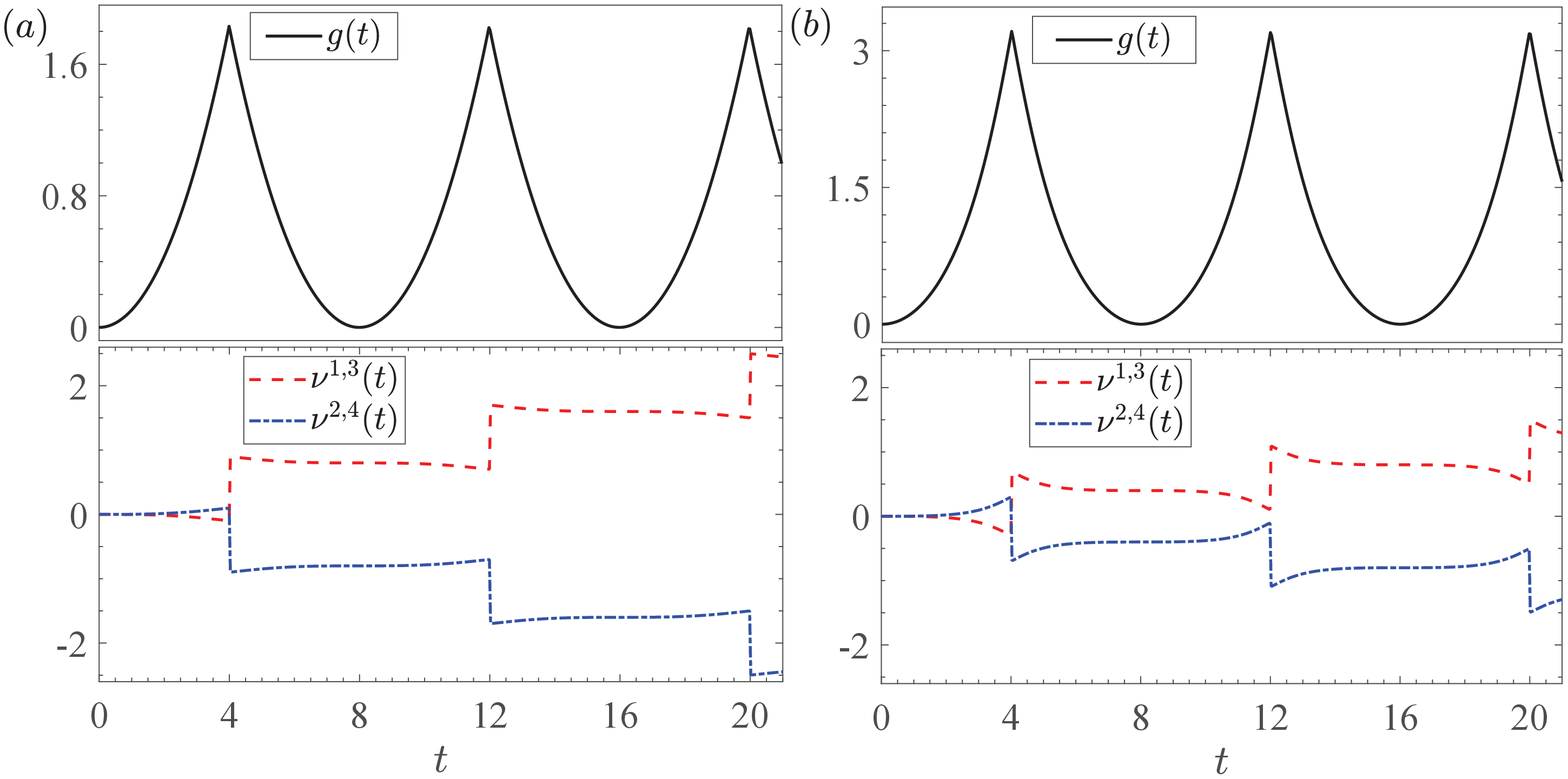}
\caption{Rate function (upper row) and DTOPs (lower row) for unitary dynamics of mixed states. The coin parameters are the same as those in Fig.~\ref{fig:mixed} of the main text.}
\end{figure}

\section{DQPT for mixed states}
We now discuss the simulation of DQPTs for mixed states, where the Loschmidt amplitude in each $k$-sector is defined as~\cite{mixeddtop}
\begin{align}
G(k,t):=\text{Tr}\left[\rho_0\left(U^\text{f}_k\right)^t\right].
\end{align}
For an initial density matrix $\rho_0=p\ket{\psi^\text{i}_-}\bra{\psi^\text{i}_-}+(1-p)\ket{\psi^\text{i}_+}\bra{\psi^\text{i}_+}$, we have
\begin{align}
G(k,t)=p(e^{iE^\text{f}_kt}|c_-|^2+e^{-iE^\text{f}_kt}|c_+|^2)
+(1-p)(e^{-iE^\text{f}_kt}|c_-|^2+e^{iE^\text{f}_kt}|c_+|^2)
:=|G(k,t)|e^{i\phi(k,t)}.
\end{align}
It is apparent that the critical momenta and critical times satisfy the same relations as in the pure-state case, with $|c_-(k_c)|=|c_+(k_c)$ and DQPTs occurring at $t_c=(2n-1)t_0$. Also similar to the pure-state case, there is only one critical-time scale $t_0=\pi/(2E^\text{f}_{k_c})$.

DTOPs in this case are defined as
$\nu^m(t)=\frac{1}{2\pi}\int^{k_{m+1}}_{k_m}\text{d}k\frac{\partial\phi^G(k,t)}{\partial k}$. However, as $\phi^G(k_m,t)\neq \phi^G(k_{n},t)$ for $m\neq n$, $\phi^G(k,t)$ is no longer periodic in the interval $k\in(k_m,k_{m+1})$, and $\nu^m(t)$ in the mixed-state case are not quantized. In Fig.~S2, we show time evolutions of the rate function and DTOP for a typical unitary dynamics of mixed state. Whereas non-analyticities still occur in $g(t)$ periodically and $\nu^m(t)$ still demonstrate abrupt jumps at critical times, $\nu^m(t)$ are not quantized.

\section{DQPTs for non-unitary dynamics}

We now discuss the simulation of DQPTs in $\mathcal{PT}$-symmetric non-unitary dynamics of pure states. As we will prove later, DQPTs occur when the system is quenched between different topological phases in the $\mathcal{PT}$-unbroken regime.

We consider the Loschmidt amplitude in each $k$-sector
\begin{align}
\tilde{G}(k,t):=\langle\psi^\text{i}_-(k)|\tilde\psi_-(t)\rangle
=b_+\tilde{c}_+e^{-i\tilde{E}^\text{f}_kt}+b_-\tilde{c}_-e^{i\tilde{E}^\text{f}_kt},
\end{align}
where $b_\pm=\langle\psi^\text{i}_-|\tilde\psi^\text{f}_\pm\rangle$, $\tilde{c}_\pm(k)=\langle\tilde{\chi}^\text{f}_\pm|\psi^\text{i}_-\rangle$. For $\tilde{U}^\text{f}$ in the $\mathcal{PT}$-unbroken regime, $\tilde{E}^\text{f}_k$ is real for all $k$.
The time-evolved state is then
\begin{align}
|\tilde\psi_-(t)\rangle=(\tilde{U}^\text{f}_k)^t\ket{\psi^\text{i}_-}
=\tilde{c}_+e^{-i\tilde{E}^\text{f}_kt}\ket{\tilde\psi^\text{f}_+}+\tilde{c}_-e^{i\tilde{E}^\text{f}_kt}\ket{\tilde\psi^\text{f}_-},
\end{align}

According to the expression of $\tilde{G}(k,t)$, DQPTs occur periodically at $t_0=(n+\frac{1}{2})\pi/\tilde{E}^{\text{f}}_{k_c}$ ($n\in \mathbb{N}$) when $\tilde{E}_{k_c}^{\text{f}}$ is real. Here $k_c$ satisfies $|b_+(k_c)\tilde{c}_+(k_c)|=|b_-(k_c)\tilde{c}_-(k_c)|$.
When the system is quenched between different topological phases in the $\mathcal{PT}$-symmetry-unbroken regime, fixed points with $\tilde{c}_+=0$ and those with $\tilde{c}_-=0$ always emerge in pairs. As $|b_+\tilde{c}_+|-|b_-\tilde{c}_-|$ are continuous functions of $k$, there must be at least one critical momentum $k_c$ satisfying $|b_+(k_c)\tilde{c}_+(k_c)|=|b_-(k_c)\tilde{c}_-(k_c)|$ inbetween two fixed points of different kinds. We thus conclude that DQPTs necessarily occur in this case. At these critical times, the rate function, defined as
\begin{align}
\tilde{g}(t)=-\frac{1}{\pi}\int^{\pi}_{-\pi}\text{d}k\ln |\tilde{G}(k,t)|
\end{align}
 becomes non-analytic. A typical behavior of $\tilde{g}(t)$ is shown in Fig.~S3(a).

\begin{figure}
\includegraphics[width=16cm]{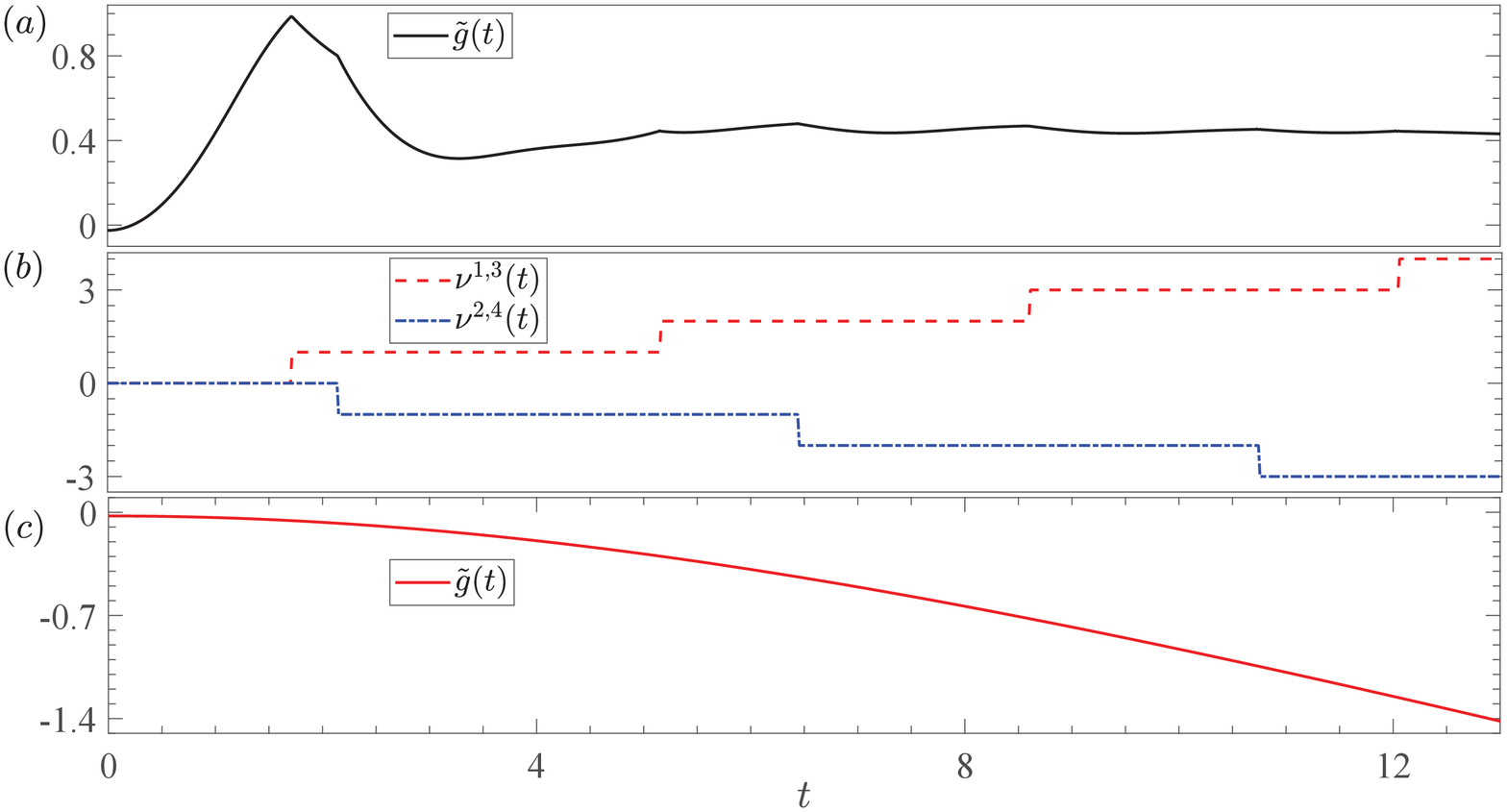}
\caption{Rate function (a) and DTOPs (b) for non-unitary dynamics in the $\mathcal{PT}$-symmetry-unbroken regime. (c) Rate function for non-unitary dynamics in the $\mathcal{PT}$-symmetry-broken regime. The coin parameters are the same as those in Fig.~\ref{fig:nonunitary} of the main text.}
\end{figure}

Similar to the unitary case, we define DTOPs between two adjacent fixed points as
\begin{equation}
  \tilde{\nu}^{m}(t)=\frac{1}{2\pi}\int_{k_{m}}^{k_{m+1}}\frac{\partial\tilde{\phi}_{k}^{G}(t)}{\partial k}\text{d}k.
\label{eqn:nuDf}
\end{equation}
Here, $\tilde{\phi}_{k}^{G}(t)=\tilde{\phi}_{k}(t)-\tilde{\phi}_{k}^{\text{dyn}}(t)$, where $\tilde{\phi}_{k}(t)$ is defined through $\tilde{G}(k,t)=|\tilde{G}(k,t)|e^{i \tilde{\phi}_{k}(t)}$ and the dynamic phase $\tilde{\phi}_{k}^{\text{dyn}}(t)=(b_-\tilde{c}_- - b_+\tilde{c}_+)\tilde{E}^\text{f}_k t$. At critical points, $\tilde{G}(k,t)$ vanishes, which leads to abrupt jumps in $\tilde{\nu}^m(t)$. Further, as $\tilde{\phi}_k^{G}(t)$ vanishes at fixed points, $\tilde{\nu}^m$ characterizes the $S^1\rightarrow S^1$ mapping from the momentum submanifold between $k_m$ and $k_{m+1}$ to $e^{i\tilde{\phi}_k^{G}(t)}$, and is therefore quantized.

Importantly, in non-unitary dynamics, due to the breaking of time-reversal symmetry, fixed points $k_m$ are no longer located at high-symmetry points. As a consequence, $\tilde{E}^\text{f}_{k_c}$ can take two different values for $k_c$ inbetween different fixed points. This gives rise to two different critical time scales [see Fig.~S3(b)].

In contrast, when the final $\tilde{U}^\text{f}$ is in the $\mathcal{PT}$-symmetry-broken regime, DQPTs are typically absent. As illustrated in Fig.~S3(c), the rate function is smooth in time, and DTOPs are not well defined due to the absence of fixed points.

\section{Fixed points and topology}

In this section, we show that the existence and number of fixed points are intimately connected with winding numbers $\nu^{\text{i},\text{f}}$ of the Floquet operators $\tilde{U}^{\text{i},\text{f}}$ respectively, when both Floquet operators belong to the $\mathcal{PT}$-symmetry-unbroken regime with completely real eigenspectra. We focus on the non-unitary case, as derivations in the unitary case with $U$ are essentially the same when taking the loss parameter $l=0$.

Starting from Eq.~(\ref{eqn:U_nonunitary}), and focus on the $\mathcal{PT}$-symmetry-unbroken regime, where $d^2_0<1$ for $\forall k$, we derive the right and left eigenvectors of $\tilde{U}_k$
\begin{align}
\ket{\tilde{\psi}_{\pm}}&=V^\dag\frac{1}{\sqrt{2\cos2\Omega}}(\pm e^{\pm i\Omega},e^{+i\vartheta}e^{\mp i\Omega})^T,\\
\bra{\tilde{\chi}_{\pm}}&=\frac{1}{\sqrt{2\cos2\Omega}}(\pm e^{\pm i\Omega},e^{-i\vartheta}e^{\mp i\Omega})V.
\end{align}
Here $V=e^{i\frac{\pi/2}{2}\sigma_y}$, $\vartheta$ and $\Omega$ are respectively defined through $-d_3+i d_2 = d e^{i\vartheta}$ and $\sin2\Omega = -id_1/d$, with $d^2=d_2^2+d_3^2$. As $d_1/d\in(0,1)$, we set $2\Omega\in(0,\pi/2)$ and $\cos2\Omega>0$. Notice that the orthonormal conditions $\left(\langle\tilde{\chi}_{\pm}|\tilde{\psi}_{\pm}\rangle = 1, \langle\tilde{\chi}_{\pm}|\tilde{\psi}_{\mp}\rangle = 0\right)$ are always satisfied in this region.

We then have
\begin{equation}
\frac{\text{d}}{\text{d}k}\ket{\tilde{\psi}_{\pm}}=V^\dag\frac{1}{(2\cos2\Omega)^{3/2}}(i2e^{\mp i\Omega}\Omega',ie^{i\vartheta}e^{\mp i\Omega}(2\cos2\Omega)\vartheta'\mp i2e^{i\vartheta}e^{\pm i\Omega}\Omega')^T,
\end{equation}
where $\Omega'=\text{d}\Omega/\text{d}k$ and $\vartheta'=\text{d}\vartheta/\text{d}k$. It is then straightforward to derive the Berry connection $A_\pm=-i\bra{\tilde{\chi}_{\pm}}\frac{\text{d}}{\text{d}k}\ket{\tilde{\psi}_{\pm}}
/\langle\tilde{\chi}_{\pm}|\tilde{\psi}_{\pm}\rangle
=\frac{1}{2}\vartheta'\pm\frac{i}{2}\vartheta'\tan2\Omega$, and hence $A_++A_-=\vartheta'$.

The winding number of $\tilde{U}$ is defined as $\nu =\varphi_\text{B}/2\pi$, where the global Berry phase $\varphi_\text{B}=\varphi_{Z+}+\varphi_{Z-}$, with
\begin{align}
\varphi_{Z\pm} =\oint \text{d}k A_\pm
=-i\oint \text{d}k\frac{\bra{\tilde{\chi}_{\pm}}\frac{\text{d}}{\text{d}k}\ket{\tilde{\psi}_{\pm}}}{
\langle\tilde{\chi}_{\pm}|\tilde{\psi}_{\pm}\rangle}.
\end{align}
We then have $\nu =\oint \text{d}k \vartheta'$. Since $\vartheta=\arg(-d_3+i d_2)=\arg(-n_3+i n_2)=\arg(i(n_2+i n_3))$, $\tilde{U}$ has the same winding number as $U$. Here the vector $\bm{n}$ is defined in Eq.~(\ref{eqn:unitaryn}).

It follows that
\begin{align}
\tilde{c}_\pm=\langle\tilde{\chi}^{\text{f}}_{\pm}|\psi^\text{i}_-\rangle=\frac{\mp e^{-i(\Omega^\text{i}\mp\Omega^{\text{f}})}+e^{i\vartheta^{0}}e^{i(\Omega^{\text{i}}\mp\Omega^{\text{f}})}}
{2\sqrt{\cos2\Omega^{\text{i}}\cos2\Omega^{\text{f}}}},\label{eqn:Cpm}
\end{align}
where $\vartheta^0=\vartheta^{\text{i}}-\vartheta^{\text{f}}$. Consider a unit vector $\bm{n}^0$ on the $y$-$z$ plane whose polar angle is given by $\vartheta^0$. As $k\in \text{1BZ}$, the number of times $\bm{n}^0$ winds around the $x$-axis is therefore $\oint \text{d}k \partial \phi^{0}/\partial k=\nu^{\text{f}}-\nu^{\text{i}}$.
From Eq.~(\ref{eqn:Cpm}), the condition for $\tilde{c}_\pm=0$ is $\phi^0=2(\Omega^\text{f}-\Omega^\text{i})$ (mod $2\pi$) and $\phi^0=\pi-2(\Omega^\text{f}+\Omega^\text{i})$ (mod $2\pi$), respectively. Therefore, the number of fixed points with $\tilde{c}_+=0$ or $\tilde{c}_-=0$ should be at least $\left|\nu^{\text{f}}-\nu^{\text{i}}\right|$ each.

Finally, we note that the existence of fixed points are not guaranteed when either $\tilde{U}^\text{i}$ or $\tilde{U}^\text{f}$ is in the $\mathcal{PT}$-symmetry-broken regime. In the absence of fixed points, topological DQPTs are absent and DTOPs cannot be defined.

\end{widetext}

\end{document}